\documentclass[%
 reprint,
superscriptaddress,
 amsmath,amssymb,
 aps,
prl,
]{revtex4-1}
\usepackage{multirow}
\usepackage{graphicx}
\usepackage{dcolumn}
\usepackage{bm}
\usepackage{todonotes}
\usepackage{simplewick}
\usepackage{physics}
\usepackage{verbatim}
\usepackage{float}
\usepackage{array}
\usepackage[utf8]{inputenc}
\usepackage[T1]{fontenc}

\usepackage{xr-hyper}
\usepackage{hyperref}
\hypersetup{
     colorlinks,
     linkcolor={blue!50!black},
     citecolor={blue!50!black},
     urlcolor={blue!80!black}
}

\usepackage{xspace}
\usepackage{xcolor}
\makeatletter
\DeclareRobustCommand\onedot{\futurelet\@let@token\@onedot}
\def\@onedot{\ifx\@let@token.\else.\null\fi\xspace}

\makeatother

\begin{document}

\title{Ab initio calculation of symmetry-breaking observables}%

\author{A. Belley}%
\thanks{Co-lead author}
\email{abelley@mit.edu}
\affiliation{Massachusetts Institute of Technology,
Cambridge, Massachusetts 02139, USA}%

\author{B. Romeo}%
\thanks{Co-lead author}
\email{bromeo@unc.edu}
\affiliation{Department of Physics and Astronomy, University of North Carolina, Chapel Hill, North Carolina 27516-3255, USA.}%

\author{J. Engel}%
\affiliation{Department of Physics and Astronomy, University of North Carolina, Chapel Hill, North Carolina 27516-3255, USA.}%

\author{D. Kekejian}
\affiliation{Quantum CodeX, 921 Woodgate Blvd. Baton Rouge, LA 70808, USA}

\author{T. Miyagi}%
\affiliation{Center for Computational Sciences, University of Tsukuba, 1-1-1 Tennodai, Tsukuba 305-8577, Japan}

\author{S. Foster}
\affiliation{Department of Engineering Physics, McMaster University, Hamilton, Ontario, L8S 4M1, Canada}
\affiliation{TRIUMF 4004 Wesbrook Mall, Vancouver BC V6T 2A3, Canada}

\author{P. Navratil}
\affiliation{TRIUMF 4004 Wesbrook Mall, Vancouver BC V6T 2A3, Canada}
\affiliation{University of Victoria, 3800 Finnerty Road, Victoria, British Columbia V8P 5C2, Canada}

\author{B. C. He}%
\affiliation{Department of Physics and Astronomy, University of Tennessee, Knoxville, TN 37996, USA}%

\author{S. R. Stroberg}%
\affiliation{Department of Physics and Astronomy, University of Notre Dame, Notre Dame, IN 46556 USA}

\author{J. D. Holt}%
\affiliation{TRIUMF 4004 Wesbrook Mall, Vancouver BC V6T 2A3, Canada}%
\affiliation{Department of Physics, McGill University, Montr\'eal, QC H3A 2T8, Canada}%

\author{R. F. Garcia Ruiz}%
\affiliation{Massachusetts Institute of Technology,
Cambridge, Massachusetts 02139, USA}%

\begin{abstract}

Symmetry-violating observables such as the nuclear anapole and Schiff moments provide sensitive probes of the fundamental symmetries of nature and physics beyond the Standard Model.  
Their interpretation has been hindered, however, by the lack of \textit{ab initio} nuclear structure calculations in the medium-mass and heavy nuclei of interest to experimentalists.   
To provide them, we introduce a new version of the in-medium similarity renormalization group (IMSRG) designed to target parity-violating operators.  
By generalizing the IMSRG flow equations to evolve the weak symmetry-breaking Hamiltonian --- and the anapole or Schiff operators --- alongside the strong nuclear Hamiltonian,  we construct a systematically improvable framework for computing these parity-violating moments. 
We benchmark the method against the no-core shell model in light nuclei and obtain the first \textit{ab initio} predictions of the anapole moment in $^{29}$Si and the Schiff moments in $^{129}$Xe. These heavier systems are of direct experimental interest.    

\end{abstract}

\maketitle

\paragraph{\textbf{\textit{Introduction.}---}}Violations of fundamental symmetries underlie some of the most sensitive tests of physics beyond the Standard Model. 
Parity-violating nuclear observables such as anapole and Schiff moments, the latter of which also violates time-reversal symmetry and induces atomic and molecular electric dipole moments (EDMs), play particularly important roles~\cite{Engel2013,Engel2025}. 
The Standard Model predicts extremely small values for these quantities, and any deviations from these predictions can indicate new physics that violates parity (P) symmetry and, in the case of Schiff moments, charge-parity (CP) symmetry, which is equivalent to time-reversal (T) symmetry.
Because stronger CP violation than that in the Standard Model is required to explain the universe's matter–antimatter asymmetry, EDM experiments are particularly compelling probes of new physics \cite{Chupp19}.
Symmetry-violating observables thus provide a low-energy window into physics that may elude collider experiments.  
They will be especially useful if we can compute their dependence on any underlying new physics. 

Unfortunately, theoretical predictions of anapole and Schiff moments remain highly uncertain, as they have relied on phenomenological models, typically based on mean-field theory and approximations to P-violating nucleon–nucleon (NN) forces~\cite{Dmitriev94,Flambaum97, Haxton02,Dmitriev05,Engel2010}.

While useful in guiding expectations, such models lack the precision required to make reliable, quantitative connections between measured quantities and the underlying symmetries of the Standard Model or its extensions.  
The interplay between strong nuclear correlations and the subtle effects of P violation is difficult to fully capture within phenomenological models, except, perhaps, when nuclear structure such as octupole deformation causes a collective enhancement.

In recent years, \textit{ab initio} theory~\cite{Hergert2020,Ekstrom2023} rooted in chiral effective field theory (EFT)~\cite{Epelbaum2009,Machleidt2011} and new many-body methods has improved our ability to predict nuclear observables.  
One of these methods, the in-medium similarity renormalization group (IMSRG)~\cite{Hergert2016, Hergert2017}, has proved to be both systematically improvable and versatile; it successfully describes ground-state energies~\cite{Hergert2016,Simonis17}, charge radii, excitation spectra, magnetic moments and transition rates, $\beta$-decay rates, and dark-matter scattering across large regions of the nuclear chart~\cite{Simonis17,Parzuchowski17, Gysbers2019,Stro21Drip,Belley2021, Yao2021,Miyagi24,Hu22SDDM}. 
The last few years have seen substantial advances in our ability to quantify the uncertainties in these calculations~\cite{Hu2022, Belley2024, Belley2025,Munoz26FRAME}. 
Yet, the IMSRG has so far been been applied only to symmetry-conserving observables, leaving parity-violating moments unexplored. 

Here, we introduce a new extension of the IMSRG designed to target symmetry-breaking observables, with an emphasis on anapole and Schiff moments.  
By generalizing the flow equations to consistently evolve P-violating operators alongside the nuclear Hamiltonian, our framework provides a path toward robust computation of these observables. 
To validate the method, we benchmark our results against no-core shell model (NCSM) calculations in light nuclei and present the first \textit{ab initio} predictions for heavier systems targeted by experiment.

\paragraph{\textbf{\textit{Methods.}---}}
P-violating nuclear observables arise from the addition of tiny symmetry-breaking interactions to the dominant P-conserving correlations.  
Because the P-violating interactions are so weak, their effects can be treated in first-order perturbation theory. 
In that approach, one typically begins with a P-conserving strong-interaction Hamiltonian $H_{PC}$, whose eigenstates $\{ | \Psi_k\rangle  \}$ form an unperturbed basis, and 
then treats the weak P-violating interaction $V_{\text{PV}}$ in first order within that basis. 
If $| \Psi_0 J^\pi \rangle$ denotes the ground state of the P-conserving Hamiltonian with angular momentum $J$ and parity $\pi$, the perturbed state can be written as
\begin{equation}
|\Psi_0 J \rangle \approx |\Psi_0 J^\pi \rangle 
+ \sum_{k \neq 0} |\Psi_k J^{-\pi} \rangle  \frac{\langle \Psi_k J^{-\pi} | V_{\text{PV}} | \Psi_0 J^\pi \rangle}{E_0 - E_k},\label{eq:fowf}
\end{equation}
where $E_0$ and the $E_k$ are the unperturbed eigenvalues.  A P-violating observable $\mathcal{O}_{\text{PV}}$ (such as the Schiff or anapole operator) acquires a nonzero expectation value through a sum over opposite-parity components of the perturbed wave function,
\begin{equation}
\langle \mathcal{O}_{\text{PV}}\rangle 
\approx 2 \, 
\sum_{k \neq 0} \frac{\langle \Psi_0 J^\pi | \mathcal{O}_{\text{PV}} | \Psi_k J^{-\pi} \rangle 
      \langle \Psi_k J^{-\pi} | V_{\text{PV}} | \Psi_0 J^\pi \rangle}
     {E_0 - E_k}.\label{eq:pvsm}
\end{equation}

Perturbation theory makes explicit the dependence of P-violating observables on the structure of excited states, through transition matrix elements of both the P-breaking interaction and the relevant operators. 
Although Eq.\ (\ref{eq:pvsm}) requires one to solve only the unperturbed Schr\"odinger equation, it often also demands access to a large number of excited-state wave functions. 
For an approach such as the IMSRG, which focuses on low-lying states in complex heavy nuclei, that access is not available without a potentially complicated extension such as the equations-of-motion method. 
The same is true of the phenomenological shell model; typical valence spaces do not contain the important intermediate states in Eq.\ \eqref{eq:pvsm}, which involve excitations of particles up to three shells above the Fermi surface.  

To overcome these limitations, we avoid the explicit construction of intermediate states in Eq.~\eqref{eq:pvsm}

by including P-violating interactions in the IMSRG flow equations. 
Density-functional theory, which involves mean-field calculations with phenomenological interactions, was extended in Ref.\ \cite{Ban2010} to treat $V_{PV}$ by simply adding it to the Hamiltonian in the mean-field equations.  
The IMSRG offers an \textit{ab initio} framework for this sort of treatment:~by evolving both the strong Hamiltonian and symmetry-breaking potential through a continuous unitary transformation, one can consistently decouple the low-energy dynamics and compute P-violating observables without requiring explicit sums over excited states. 
In the following, we outline the extension of the IMSRG that enables this treatment, a method we call the Parity-Violating IMSRG (PV-IMSRG).  

In its standard formulation, the IMSRG employs a continuous unitary transformation to reshape the Hamiltonian into a form that is easier to diagonalize. 
It obtains the transformation through the flow equation,
\begin{equation}\label{eq:flow}
    \frac{dH(s)}{ds} = \left[\eta(s), H(s)\right],
\end{equation}
where $H(s)$ is the Hamiltonian as a function of the flow parameter $s$, and $\eta(s)$, the generator of the transformation, encodes the degrees of freedom one wishes to decouple. 
Any other operator $O$ must be evolved consistently under the same transformation via the equation
\begin{equation}
    \frac{dO(s)}{ds} = \left[\eta(s), O(s)\right].
\end{equation}

To use this framework for our purposes, we start from a Hamiltonian that includes both P-conserving ($H_{PC}$) and P-violating ($V_{PV}$) interactions,
\begin{equation}
    H = H_{PC} + \lambda V_{PV},\label{eq:H}
\end{equation}

where $\lambda$ is a power-counting parameter, ultimately set to 1, used to track small terms. 
The corresponding generator can be written as
\begin{equation}
    \eta = \eta_{PC} + \lambda \eta_{PV}.\label{eq:eta}
\end{equation}
Inserting these definitions into Eq.~\eqref{eq:flow}, we obtain the coupled flow equations, 
\begin{align}
    \frac{dH_{PC}(s)}{ds} &= \left[\eta_{PC}(s), H_{PC}(s)\right] 
    + \lambda^2 \left[\eta_{PV}(s), V_{PV}(s)\right],\label{eq:HPC_evol} \\
   \frac{dV_{PV}(s)}{ds} &= \left[\eta_{PC}(s), V_{PV}(s)\right]+ \left[\eta_{PV}(s), H_{PC}(s)\right]\label{eq:HPV_evol}.
\end{align}
Because $V_{PV}$ is so weak, we can neglect the $\mathcal{O}(\lambda^2)$ term in Eq.\ (\ref{eq:HPC_evol}), so that $V_{PV}$ does not affect the flow of $H_{PC}$.
For a generic operator $O = O_{PC} + O_{PV}$ consisting of P-conserving and P-violating parts, we obtain the similar equations,
\begin{align}
    \frac{dO_{PC}(s)}{ds} &= \left[\eta_{PC}(s), O_{PC}(s)\right] 
    + \lambda \left[\eta_{PV}(s), O_{PV}(s)\right],\label{eq:OPC_evol} \\
    \frac{dO_{PV}(s)}{ds} &= \left[\eta_{PC}(s), O_{PV}(s)\right] + \lambda \left[\eta_{PV}(s), O_{PC}(s)\right]\label{eq:OPV_evol}.
\end{align}

An important feature of these equations is that even an operator that initially has negative parity, so that $O_{PC}(0)=0$, will acquire a P-conserving part that represents, in the Heisenberg picture, the effects of opposite-parity virtual excitations.   
If the states in the low-energy space that we decouple all have the same parity, then that positive-parity operator is all that matters for those states at the end of the flow.

The Magnus formulation of the IMSRG~\cite{Morris2015} provides an alternative view of the method.
First, we transform all operators (i.e. integrate to $s\to\infty$) using Eqs.~\eqref{eq:HPC_evol}-\eqref{eq:OPV_evol} with $\eta_{PV}=0$.
Then, retaining terms of order $\lambda$, we can effectively invert the propagator in Eq.~\eqref{eq:pvsm} via the decoupling condition 
\begin{equation}\label{eq:magnusPV}
    [H_{PC},\Omega_{PV}]=V_{PV} \,,
\end{equation}
where $e^{\Omega_{PV}}$ is the parity-violating unitary transformation.

In terms of this operator, the desired moment is
\begin{equation}\label{eq:magnusOPV}
    \langle \mathcal{O}_{PV}\rangle = \langle \Psi_0 J^\pi | \,[\Omega_{PV},\mathcal{O}_{PV}] \,| \Psi_0 J^\pi \rangle \,.
\end{equation}

Equation~\eqref{eq:magnusPV} can be solved for $\Omega_{PV}$ by iteration, or by integrating a flow equation.
In general, $\Omega_{PV}$ can include 3-body or higher-body pieces, which we discard in the IMSRG(2) approximation.
When operators are truncated at the 2-body level, the present approach is equivalent to equations-of-motion with up to 2-particle-2-hole excitations.

In this paper, we apply these ideas within the valence-space formulation of the IMSRG (VS-IMSRG)~\cite{Stro19ARNPS} as implemented in the \texttt{imsrg++} code~\cite{Stroberg_IMSRG_2018}. 
Within that approach, the flow equations lead to an effective Hamiltonian that acts within a pre-selected valence space, with eigenvectors 
that lie fully in that space. 
To diagonalize the effective Hamiltonian, we use the large-scale shell-model code \texttt{Kshell}~\cite{shimizu2019thick}. 
We employ the IMSRG(2) approximation, in which operators are normal ordered with respect to a reference state and subsequently truncated at the two-body level. 
This formulation enables the treatment of many of the medium-mass and heavy nuclei used in measurements of P-violating observables.

\paragraph{\textbf{\textit{The anapole moment.}---}} The anapole moment is a P-violating, time-reversal-conserving nuclear moment caused by weak corrections to electromagnetic currents~\cite{Zeldovich57,Zeldovich57b}. 
It can thus be used both to constrain our description of the hadronic weak interaction and to search for particles such as a new $Z'$ boson and other dark matter candidates~\cite{Alves14,Alves18,Safronova2018,Gaul2025}.  
For either task, accurate nuclear structure calculations of the anapole moment are essential.

Here we focus on the dominant contribution to the anapole moment, which arises from the  nuclear spin~\cite{Haxton89}. 
The corresponding operator is 
\begin{equation}
    \boldsymbol{a} = \frac{\pi e}{m}\sum_{i=1}^A \mu_i (\boldsymbol{r}_i \times \boldsymbol{\sigma}_i),
\end{equation}
where $e$ is the electron charge, $m$ the nucleon mass, and $\mu_i$ the magnetic moment of the $i^\text{th}$ nucleon (neutron or proton) in units of nuclear magnetons. 
For the P-violating interaction $V_{PV}$ we use the Desplanques, Donoghue, and Holstein NN potential~\cite{Desplanques1980}, with updated parameter values from Ref.~\cite{Blyth2018}. 
This choice allows us to compare our results with those from  Ref.~\cite{Hao2020}, obtained in the NCSM in light nuclei. 
Potentials based on chiral effective field theory have been derived~\cite{deVries20}, and we will implement them in the future.

The size of P-violating atomic effects induced by the anapole moment is proportional to the quantity 
\begin{equation}
    \kappa_A = \frac{\sqrt{2}e}{G_F}a \,,
\end{equation}
where $G_F$ is the Fermi constant and
\begin{equation}
\label{eq:anapole-defn}
    a = \langle \Psi_0 J, M=J| a_{z} | \Psi_0 J, M=J\rangle \,.
\end{equation}
In the absence of reliable many-body calculations of the anapole moment, rough ``single-particle'' estimates are commonly used to guide and motivate experiments. 
We refer the reader to the Appendix and references therein for a more detailed discussion of them. 

\paragraph{\textbf{\textit{The Schiff moment.}---}}
When a neutral atom is placed in a uniform electric field, the electrons rearrange themselves so that there is almost no net electric field in the nucleus, and any nuclear electric dipole moment due to CP-violating physics is shielded. 
As Schiff originally showed~\cite{Schiff63}, this shielding effect is perfect in the limit of a point-like nucleus and non-relativistic electrons~\cite{Engel2000} . The finite-sized charge distribution of the nucleus mitigates this screening, however, and as a result, CP-violating (and P- and T-violating) nuclear effects become observable through higher-order moments of the charge distribution, with the leading contribution in many cases arising from the nuclear ``Schiff moment.''  This quantity, to which electric dipole moments in diamagnetic atoms and molecules containing them  are proportional, is given by 
\begin{equation}
    S=\bra{\Psi_0 J,M=J}S_{z}\ket{\Psi_0 J,M=J} \,,\label{eq:SM-defn}
\end{equation}
where the most important part of the Schiff operator $\mathbf{S}$, reflecting a P- and T-violating charge distribution, is 
\begin{equation}
    \mathbf{S}=\frac{e}{10}\sum_{i=1}^Z\left[r_i^2-\frac{5}{3}\langle r^2\rangle_{\rm ch}\right]\mathbf{r}_i\label{Schiff_pl_op} \,.
\end{equation}
Here $\langle r^2\rangle_{\rm ch}$ is the nuclear charge radius, $Z$ is the proton number, and $\mathbf{r}_i$ the position vector of the $i^\text{th}$ proton relative to the center of mass.

The Schiff moment is generated by the modification of the nuclear charge distribution produced by P- and T-violating nucleon–nucleon force, the dominant part of which at leading order in $\chi$EFT is given by the one-pion exchange potential introduced in Refs.~\cite{Haxton37,Roberson87,Liu04,deVries20}. 
The potential is proportional to the strong $\pi NN$ coupling constant $g\approx 13.5$ and three dimensionless constants $\bar{g}^{(0)}_{\pi}$, $\bar{g}^{(1)}_{\pi}$, and $\bar{g}^{(2)}_{\pi}$, that correspond to P- and T-violating  isoscalar, isosvector, and isotensor couplings induced by new CP-violating physics. 

Like the anapole moment in Eq.~\eqref{eq:anapole-defn}, the Schiff moment is defined as the expectation value of the $z$-component of the corresponding operator in the state with the largest spin projection. The definitions in both Eq.~\eqref{eq:anapole-defn} and \eqref{eq:SM-defn} allow a direct comparison to experimental observables in fully polarized nuclei.

The weakness of the P- and T-violating potential means that the $S$ depends essentially linearly on the couplings $\bar{g}_\pi^{i}$, and one can write it in the form
\begin{equation}
S=a_0 g\bar{g}^{(0)}_{\pi}
+a_1 g\bar{g}^{(1)}_{\pi} +a_2 g\bar{g}^{(2)}_{\pi} \,,
\label{eq:ai_coeffs}
\end{equation}
where the $a_i$, the coefficients in the linear relation, are what must be computed. 
The Schiff operator can raise or lower single-particle energies by up to three oscillator quanta, so no realistic valence space will be large enough to contain all the orbitals that are important for excited states. 
Something like our procedure is required.

The literature contains a simple single-particle estimate of Schiff moments~\cite{Flambaum86} as well as anapole moments~\cite{Sushkov84}.
The former are more strongly affected by collectivity, however, and so the single-particle estimate is not often used nowadays. 
We again refer the reader to the Appendix for more details. 

\paragraph{\textbf{\textit{Benchmark in light isotopes.}---}} 

Although the IMSRG(2) approximation has been tested extensively with the usual P-conserving interactions~\cite{Stroberg2024,He2024,Heinz2021,Heinz2026}, we must verify that it does not introduce significant error in P-violating observables. 
We therefore benchmark the predictions for the anapole and Schiff moments against those of the \textit{ab initio} NCSM~\cite{Hao2020,Froese21,Foster25,Ng2026}. 

First, we confirm that the similarity transformation implemented through Eq.~\eqref{eq:eta} does indeed suppress excitations out of the valence space. 
To do so, we compute the anapole moment of \textsuperscript{2}H, 
with its two parity contributions calculated separately along the flow. 
In this system, the IMSRG(2) is exact, allowing us to check our implementation. 
We limit the calculation to the $sp$ shell where the perturbative sum in Eq.~\eqref{eq:pvsm} can be computed exactly. 

\begin{figure}[h]
    \centering
    \includegraphics[width=1\linewidth]{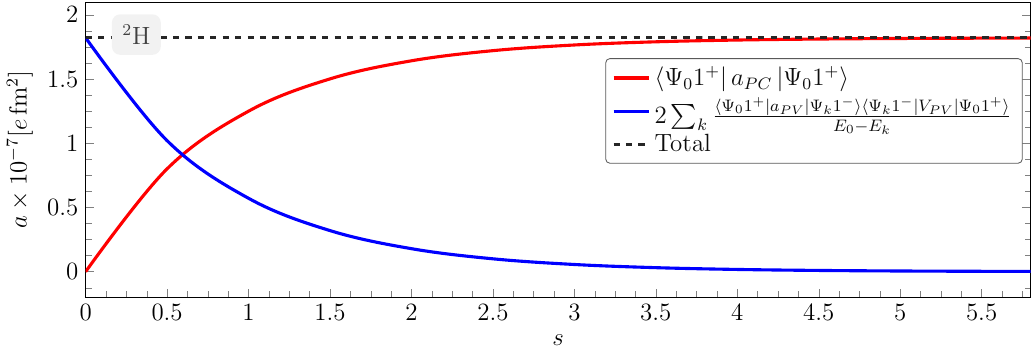}
    \caption{PV-IMSRG evolution of $a$ in \textsuperscript{2}H, broken into contributions from the direct expectation-value of the induced parity-conserving anapole-moment operator ($a_{PC}$) and the perturbative expectation value of the parity-violating operator ($a_{PV}$), and parametrized by flow parameter $s$. 
    The sum of the two contributions is always conserved, with the parity-conserving operator accounting for all of $a$ at the end.
    }
    \label{fig:FlowEvolution}
\end{figure}

Figure~\ref{fig:FlowEvolution} shows the evolution of the anapole moment along the IMSRG flow in Eq's~\eqref{eq:HPC_evol}--\eqref{eq:OPV_evol} as a function of the flow parameter $s$. 
As we perform the evolution, the moment is transferred from the expectation value of the parity-violating part of the effective anapole operator, evaluated in perturbation theory, to the expectation value of the induced parity-conserving operator in a way that preserves the sum of the contributions.
This result shows that the transformation is indeed unitary when no truncations are made. 

\begin{figure*}
    \centering
    \includegraphics[width=1.\linewidth]{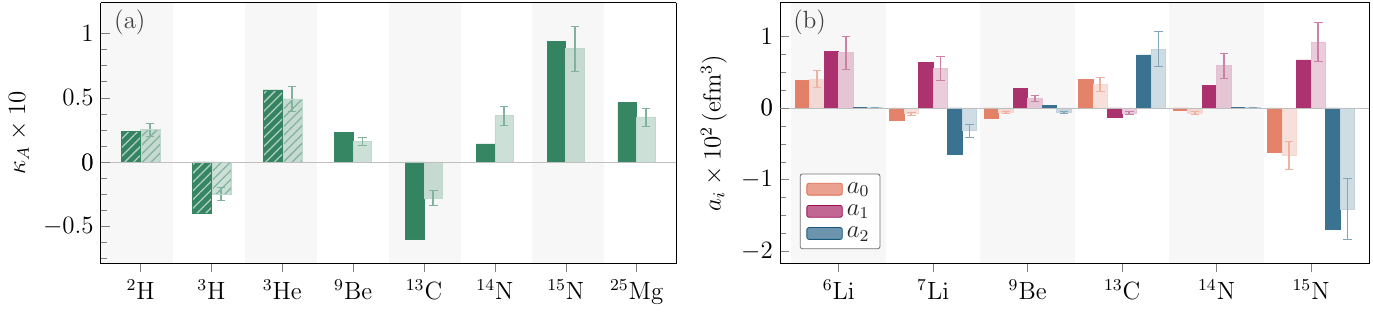}
    \caption{Comparison of (a) $\kappa_A$ and (b) 
    Schiff-moment coefficients $a_i$ obtained with PV-IMSRG (darker bars) and the NCSM (lighter bars). 
    The hashes on the first three bars in (a) mean that the corresponding values of $\kappa_a$ are multiplied by an additional factor of 10 (beyond that indicated by the $y$-axis label).
    We conservatively estimate the uncertainty in the NCSM values for the anapole and Schiff moments to be 20\% and 30\%,  respectively. 
    } 
    \label{fig:benchmark}
\end{figure*}

To compare with prior NCSM results~\cite{Hao2020,Froese21,Foster25}, we use the chiral interaction NN+3N(lnl)~\cite{Soma20}, consisting of an NN interaction from up to the fourth order ($\rm N^3$LO) in the chiral expansion~\cite{Entem03} and a $\rm N^2$LO 3N interaction with both local and non-local regulators.  Although its low energy constants are fit only to nuclei with $A=2$, 3, and 4, the interaction's description of light and medium-mass nuclei is encouraging~\cite{Soma20}.  We truncate the initial harmonic oscillator single-particle basis to states with $2n+l\leq e_{\mathrm{max}}$, where $n$ and $l$ are the radial and orbital angular momentum quantum numbers.
For all calculations we take the oscillator frequency to be $\hbar w=16$ MeV. 
We also restrict the three-body matrix elements to be non-zero only for three-particle states with $(2n_1+l_1+2n_2+l_2+2n_3+l_3)\leq E_{3\mathrm{max}}=24$, which is sufficient for the systems studied here~\cite{Miyagi2022}.

Figure~\ref{fig:benchmark} (a) shows the results for $\kappa_A$ obtained with the PV-IMSRG and the NCSM. 
In the PV-IMSRG we use $e_{\rm max}=10$, by which point the results have converged with respect to increases in model-space size. 
The PV-IMSRG moments are generally larger than those from the NCSM, a consequence mainly of the IMSRG(2) approximation. 
The differences in results, almost always less than a factor of two, are still smaller than typical discrepancies between phenomenological methods~\cite{Dmitriev97,Haxton02}.  The single-particle estimates usually agree with the complete results to within an order of magnitude, which confirms their reliability as a rough estimate, but  rarely agree as well as the results of the two \textit{ab initio} calculations agree with one another. 

These results are encouraging. IMSRG-based methods are known to face challenges in light nuclei with many-body effects such as clustering that are less pronounced in heavier isotopes. 
Nonetheless, it will be important to move beyond the IMSRG(2) level in the future.

\begin{figure}[h!]
    \centering
    \includegraphics[width=0.97\linewidth]{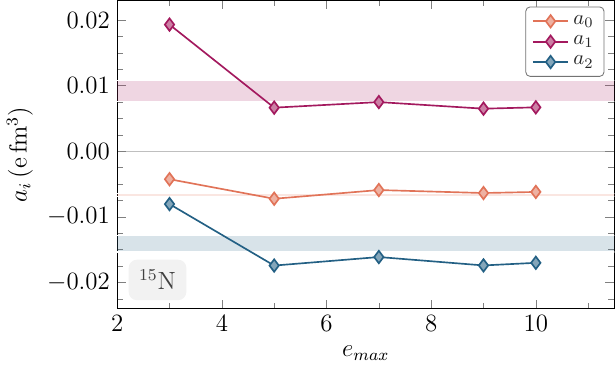}
    \caption{Schiff-moment coefficients $a_i$ in $^{15}$N (in $e\,\rm fm^3$). 
    as a function of $e_{\rm max}$, together with the NCSM values (bands). The NCSM uncertainty is from the difference in results with the largest and third largest values of $N_{\rm max}$. 
    }
    \label{fig:SM_N15_emax_cnvg}
\end{figure}

Figure~\ref{fig:benchmark} (b) shows similar results for the Schiff-moment coefficients introduced in Eq.~\eqref{eq:ai_coeffs}. 
We have again used $e_{\rm max}=10$ to obtain the PV-IMSRG numbers. Though the Schiff moment is more sensitive to admixtures of high-lying orbitals than the anapole moment, 
Fig.~\ref{fig:SM_N15_emax_cnvg} shows that in $^{15}$N, $e_{\rm max}=10$ is sufficient for convergence.  
As with the anapole moments, the disagreement between the PV-IMSRG and the NCSM in Fig.~\ref{fig:benchmark} (b) is almost never greater than a factor of two, except when the numbers are extremely small.
The largest discrepancies for unsuppressed coefficients are in $^9$Be, which displays clustering, and $^{14}$N.  
Some of the inaccuracy may also be in the NCSM, for which the particle-hole truncation level may not be sufficient, rather than in the PV-IMSRG. 

To highlight the capacity of this method to reach isotopes of interest for ongoing and future searches, we finally compute $\kappa_A$ for $^{29}$Si, which is the main target isotope in a new experimental program~\cite{Karthein2024}, and the Schiff coefficients $a_i$ in $^{129}$Xe, the lightest nucleus for which the Schiff moment has been bounded~\cite{Rosenberry01,Sachdeva2019,Allmendinger19}. 
At $e_{\rm max} = 10$, we find that
\begin{equation}
\label{eq:si29}
    \kappa_A ({^{29}\rm Si}) =  7.7 \times 10^{-2},
\end{equation}
which is several times larger than the single-particle estimate of $\kappa_A=1.2(8)\times 10^{-2}$. For the Schiff moment of $^{129}$Xe at $e_{\rm max}=14$, we obtain 
\begin{equation}
    S \left(^{129} {\rm Xe}\right) = -0.021g\bar{g}^{(0)}_{\pi}-0.010g\bar{g}^{(1)}_{\pi}-0.044g\bar{g}^{(2)}_{\pi}.
\label{eq:xe129}
\end{equation}
These numbers are reasonably close to, but always smaller than, those of the large-scale shell model~\cite{Yanase20}, ( $S=-0.038g\bar{g}^{(0)}_{\pi} -0.041g\bar{g}^{(1)}_{\pi} -0.081g\bar{g}^{(2)}_\pi$), and larger than those of the phenomenological random phase approximation~\cite{Dmitriev05}. 

The results in Eqs.\ \eqref{eq:si29} and \eqref{eq:xe129} are preliminary, with only one interaction and no variation of model parameters, and we will need careful analysis in future work to assess uncertainties in these moments. 
There are no obvious obstacles to that kind of assessment, however, both in these nuclei and in heavier isotopes such as $^{199}$Hg, which has the best experimental limit on an atomic electric dipole moment.  

\paragraph{\textbf{\textit{Conclusion.}---}}
We have presented the parity-violating IMSRG, the first \textit{ab initio} nuclear many-body method tailored to compute P- and CP-violating observables in heavy isotopes. 
The method uses the IMSRG flow equations to decouple perturbative contributions from most virtual opposite-parity excited states, with most of the relevant physics moved to an induced parity-conserving operator. 
After the flow, calculations can then be carried out in a manageable valence space. 
We benchmarked the method against the quasi-exact NCSM in light nuclei, where we find good agreement. 
We then computed the anapole moment of $^{29}$Si and the Schiff moment of $^{129}$Xe to demonstrate our ability to compute such moments in nuclei that are used in ongoing experiments. 
 Our developments pave the way for precise calculations of nuclear quantities that are essential for guiding and interpreting precision tests of fundamental-symmetry violation and searches for physics beyond the Standard Model. Future work is needed to quantify the uncertainties associated with model-space and flow-equation truncation.

\begin{acknowledgments}

The VS-IMSRG code employed in this work utilizes the Armadillo C$++$ library~\cite{Armadillo,Sanderson2016}. 
The nuclear interaction and the P-, and PT- violating potentials used in this work were generated using the \texttt{NuHamil} code~\cite{Miyagi2023}.
 Computational resources were provided by subMIT at MIT Physics, and by Sycamore and Longleaf compute clusters, administered by UNC-Chapel Hill ITS Research Computing. This work was supported by the Office of Nuclear Physics, U.S. Department of Energy, under Grant No. DE-FG02-97ER41019. 
 A.B. acknowledges the support of the Natural Sciences and Engineering Research Council of Canada (NSERC) [PDF-587464-2024].
 S.R.S. acknowledges support from the NSF under Grant No. PHY-2340834, the NSF FRHTP PHY-2402275, and the DOE Topical Collaboration “Nuclear Theory for New Physics,” Award No. DE-SC0023663.
 T.M. acknowledges support from the JST ERATO Grant No. JPMJER2304, Japan, and from JSPS KAKENHI Grant Numbers 25K07294, 25K00995, and 25K07330.
 B.C.H. acknowledges support from the National Science Foundation (NSF) FRHTP program under award No.  PHY-2402275.
J.D.H. acknowledges support from 
NSERC under grant SAPIN-2024-0003, and the Arthur B. McDonald Canadian Astroparticle Physics Research Institute.
P.N. acknowledges support from NSERC Grant No. SAPIN-2022-00019 and computing support from an INCITE Award on the Frontier supercomputer of the Oak Ridge Leadership Computing Facility (OLCF) at ORNL and from the Digital Research Alliance of Canada.
TRIUMF receives federal funding via a contribution agreement with the National Research Council of Canada.
\end{acknowledgments}


\bibliographystyle{apsrev4-1}
\bibliography{library}{}

\begin{center}
\textbf{End of Matter}
\end{center}

\begin{table}[t]
    \centering
    \scalebox{0.85}{
    \def\arraystretch{1.3}
    \begin{tabular}{cccc}
        \hline\hline
        Nucl. & PV-IMSRG & NCSM & s.p. \\
        \hline 
        $^2$H    & $0.0024 $  & $0.0025 $  &  --- \\
        $^3$H     & $-0.0040 $ & $-0.0025 $ & $-0.015(4) $ \\
        $^3$He    & $0.0056 $ &$ 0.0049 $ & $0.0027(2) $\\
        \hline
        $^9$Be    & $0.023$ & $0.016 $ & $0.007(5) $  \\
        $^{13}$C  & $-0.060 $ & $-0.028 $&  $-0.007(5) $ \\
        $^{14}$N  & $0.014 $ & $0.036 $ & $0.035(10) $ \\
        $^{15}$N  & $0.094 $ & $0.088 $ & $0.044(10) $\\
        $^{25}$Mg & $0.046 $ & $0.035 $ &  $0.015(10) $\\
        \hline\hline
    \end{tabular}} 
    \caption{Values of $\kappa_A$ obtained with the PV-IMSRG and the NCSM, and single particle estimates~\cite{Flambaum1984,Flambaum1985}. 
    The NCSM results for nuclei heavier than $^{3}$He are from Ref.~\cite{Hao2020}. Uncertainty in the single particle estimates reflects that in the coupling constants $g_i$ from Ref.~\cite{Fadeev2019}. 
    }
    \label{tab:anapole}
\end{table}

\paragraph{\textbf{\textit{Single particle estimates}---}} 
Well-known ``single-particle'' estimates, coming from the assumption that the parity violating nuclear moment is due only to a valence nucleon, are based on a picture in which both the strong nucleon-nucleon interaction and the P- or T-violating interactions are represented by mean-field potentials, and on several further approximations. 
These approximations neglect effects such as core polarization and configuration mixing that are known to be important for both the anapole and the Schiff moment. 

\begin{table}[t]
    \centering
    \scalebox{0.9}{
\def\arraystretch{1.5}
    \begin{tabular}{lccccc}
        \hline\hline
         Nucl. & & PV-IMSRG & NCSM & s.p. \\
        \hline 
        \multirow{3}{*}{$^6$Li}    &  $a_0$  &  0.0038  &  0.0040  &  0  \\
                                   &  $a_1$  &  0.0079  &  0.0077  &  0.0066 \\
                                   &  $a_2$  &  0.00005 &  0.00001 &  0  \\
        \hline                                  
        \multirow{3}{*}{$^7$Li}    &  $a_0$  &  -0.0017 & -0.0008 & -0.0010  \\
                                   &  $a_1$  &   0.0063 &  0.0055 &  0.0073  \\
                                   &  $a_2$  &  -0.0065 & -0.0032 & -0.0021  \\
        \hline                                  
        \multirow{3}{*}{$^9$Be}    &  $a_0$  &  -0.0014 &  -0.0006 & -0.0010 \\
                                   &  $a_1$  &   0.0027 &   0.0014 &  0.0086 \\
                                   &  $a_2$  &   0.0003 &  -0.0006 & -0.0019 \\
        \hline                                  
        \multirow{3}{*}{$^{13}$C}  &  $a_0$  &  0.0040  &  0.0033 &   0.0028  \\
                                   &  $a_1$  &  -0.0013 & -0.0007 &  -0.0368  \\
                                   &  $a_2$  &  0.0074  & 0.0082  &   0.0057  \\ 
        \hline                                  
        \multirow{3}{*}{$^{14}$N}  &  $a_0$  &  -0.0004 & -0.0007 &   0  \\
                                   &  $a_1$  &   0.0031 &  0.0059 &  -0.0039 \\
                                   &  $a_2$  &   0.0001 &  0.0001 &   0 \\ 
        \hline                                  
        \multirow{3}{*}{$^{15}$N}  &  $a_0$  &  -0.0062 & -0.0066  &  0.0027  \\
                                   &  $a_1$  &   0.0067 &  0.0092  & -0.0405  \\
                                   &  $a_2$  &  -0.0170 & -0.0141  &  0.0054  \\
        \hline\hline
    \end{tabular}} 
    \caption{Coefficients $a_i$ from Eq.~\eqref{eq:ai_coeffs}, calculated with PV-IMSRG and NCSM, alongside single-particle estimates~\cite{Sushkov84,Flambaum86,Engel2025},  in units of $e~\rm fm^3$. 
    NCSM calculations employ an
    oscillator frequency $\hbar \omega=20\rm MeV$, except in $^6$Li where they use $\hbar \omega=16\rm MeV$. 
    The PV-IMSRG results are with $\hbar \omega =16\rm MeV$. 
    Ref.\ \cite{Froese21} shows that the converged results for electric dipole moments do not depend $\hbar\omega$.
    } \label{tab:SM_ais}
\end{table}

The single-particle estimate for the anapole moment is given by ~\cite{Flambaum1984,Flambaum1985}
\begin{align}\label{eq:sp_anapole}
    \kappa_A &= \frac{9}{10}\frac{\alpha \mu_i}{m_p r_0}g_i A^{2/3}\frac{(J+\frac{1}{2})(-1)^{J-l_i+1/2}}{J+1}\\
    &\simeq 1.5\times 10^{-3}g_i \mu_i A^{2/3}\frac{(J+\frac{1}{2})(-1)^{J-l_i+1/2}}{J+1} \,,
\end{align}
where $\alpha \simeq 1/137$ is the fine-structure constant, $\mu_i$ is the magnetic moment of the unpaired nucleon, $m_p$ is the mass of the proton, $r_0\equiv 1.2$ fm is the scale of the nuclear radius, $A$ is the mass number, $J$ is the nuclear angular momentum , and $l_i$ the orbital angular momentum of the unpaired nucleon. 
The dimensionless constants $g_i$ are derived from the mean nucleon-nucleus weak potential for the unpaired nucleon. 
These constants are defined only roughly; here we use the values from Ref.~\cite{Fadeev2019} of $g_p = 3.4\pm0.8$ and $g_n = 0.9\pm0.6$.

For the Schiff moment, the simple model leads to the single-particle estimate~\cite{Sushkov84,Flambaum86,Engel2025}
\begin{equation}
    s_v\approx|e|\frac{1\pm (j+\frac{1}{2})}{j+1}A^{2/3}10^{-2}\varepsilon\,\rm fm^3,\label{eq:SM_sp}
\end{equation}
where $\pm$ here correspond to $l=j\pm1/2$ for the unpaired nucleon, and  
\begin{equation}
    \varepsilon=\frac{g}{2}\Bigg[\left(\frac{N-Z}{A}\right)(\bar{g}^{(0)}_\pi+2\bar{g}^{(2)}_\pi)-\bar{g}^{(1)}_\pi \Bigg].
\end{equation}

\paragraph{\textbf{\textit{Results}---}} The tables in this section present the results in Fig.~\ref{fig:benchmark} and the corresponding single-particle estimates.

\end{document}